\documentclass[11pt,a4paper]{article}
\usepackage{jstyle}

\usepackage[utf8]{inputenc}
\usepackage{amsthm,amsmath,latexsym,amssymb,amsfonts,amssymb,amscd}
\usepackage{hyperref}
\usepackage{cancel}
\usepackage{color}
\usepackage{mathabx}
\usepackage{framed,pifont}

\definecolor{rougef}{rgb}{0.56,0,0}
\definecolor{vertf}{rgb}{0,0.5,0}
\definecolor{bleuf}{rgb}{0,0,0.8}
\definecolor{violetf}{rgb}{0.5,0,0.5}

\author[1]{\quad Pan KESSEL}
\author[2]{\quad Karapet MKRTCHYAN}

\affiliation[1]{Machine Learning Group, Technische Universit\"at Berlin \\
Marchstrasse 23, 10587 Berlin, Germany}

\affiliation[2]{Max Planck Institute for Gravitational Physics (Albert Einstein Institute)\\
Am M\"uhlenberg 1, 14476 Potsdam, Germany}

\emailAdd{pan.kessel@gmail.com}
\emailAdd{karapet.mkrtchyan@aei.mpg.de}

\title{\centering
\LARGE{Cubic interactions of massless bosonic fields in three dimensions II: Parity-odd and Chern-Simons vertices}}

\abstract{This work completes the classification of the cubic vertices for arbitrary spin massless bosons in three dimensions started in a previous companion paper by constructing parity-odd vertices. Similarly to the parity-even case, there is a unique parity-odd vertex for any given triple $s_1\geq s_2\geq s_3\geq 2$ of massless bosons if the triangle inequalities are satisfied ($s_1<s_2+s_3$) and none otherwise. These vertices involve two (three) derivatives for odd (even) values of the sum $s_1+s_2+s_3$. A non-trivial relation between parity-even and parity-odd vertices is found. Similarly to the parity-even case, the scalar and Maxwell matter can couple to higher spins through current couplings with higher derivatives.
We comment on possible lessons for 2d CFT.
We also derive both parity-even and parity-odd vertices with Chern-Simons fields and comment on the analogous classification in two dimensions.}

\begin{document}

\maketitle
\section{Introduction}

The present paper is meant to complete the program set out in \cite{Mkrtchyan:2017ixk}, concerning the classification of cubic interactions for massless bosons in three space-time dimensions. In perspective, the hope is that this work will lead to a full non-linear action formulation for higher-spin systems coupled with matter in $d=3$.

Higher-Spin (HS) Gravity \cite{Vasiliev:1990en,Prokushkin:1998bq} is one of the promising attempts for the reconciliation of quantum theory and General Relativity. Conjectured dualities with known CFT's (see \cite{Gaberdiel:2010pz,Beccaria:2016tqy,Giombi:2016pvg,Bae:2017fcs} and references therein) put various models of HS gravity in the front line of holographic studies of quantum gravity. A simple and promising example of holographic duality is the $AdS_3/CFT_2$ conjecture of \cite{Gaberdiel:2010pz}.
One of the main drawbacks of these models is, however, the lack of a bulk description suitable for quantisation. This problem in particular is attacked through the so-called Fronsdal program --- a perturbative construction of classical action for HS gravity models by applying the Noether method to the gauge symmetries of massless HS fields. The starting point is the free Fronsdal action for massless fields of any spin \cite{Fronsdal:1978rb}. The interacting theories can be constructed order-by-order in powers of the fields, starting from the first non-trivial order --- cubic vertices. The latter are the main building blocks of most of the known interacting theories. 

Cubic interactions of massless higher-spin fields in arbitrary space-time dimensions $d\geq 4$ were studied extensively starting from the pioneering work \cite{Bengtsson:1983pd} later extended to the complete light-cone gauge classification of vertices first in four dimensions \cite{Bengtsson:1986kh} and then in arbitrary dimensions $d\geq 4$ \cite{Metsaev:2005ar}. The covariant approach has been developed more slowly compared to the light-cone approach, and after seminal works of the same period \cite{Berends:1984rq, Fradkin:1987ks}, the Fronsdal program \cite{Fronsdal:1978rb} was revived again in the current millenium (see  \cite{Bekaert:2006us,fms1, Fotopoulos:2008ka, Zinoviev:2008ck, Boulanger:2008tg, Manvelyan:2009tf, Bekaert:2009ud, Manvelyan:2010wp} and references therein) resulting in the classification of parity-even cubic vertices in Minkowski space of any dimension $d\geq 4$ \cite{Manvelyan:2010jr}, i.e. the covariant extension of \cite{Metsaev:2005ar}. These vertices were packed into surprisingly compact generating functions \cite{Sagnotti:2010at, Fotopoulos:2010ay, Manvelyan:2010je, Mkrtchyan:2010pp}, with intriguing hints on possible relations with String Theory, and the studies of their $(A)dS$ extensions followed \cite{Joung:2011ww,Manvelyan:2012ww,Francia:2016weg} in parallel with Vasiliev's frame-like approach \cite{Zinoviev:2010cr,Vasilev:2011xf,Boulanger:2012dx} to $(A)dS$ vertices.

Even though the light-cone classification has been known since long time, the full covariant classification in four dimensions was completed only recently in \cite{Conde:2016izb}, where the parity-odd vertices in $d=4$ were derived. A notable difference between covariant and light-cone vertices in four dimensions is the existence of a two-derivative ``minimal'' coupling to gravity in the light-cone, which is absent in the covariant classification. It is tempting to speculate that in four dimensions symmetric tensor fields may not constitute a perfect choice for minimal covariant variables for describing flat space theories and possibly even for the $(A)dS_4$ Vasiliev theory\footnote{The possibility of describing the same spectrum of particles with alternative choice of ``minimal variables'', i.e. mixed-symmetry tensors, are poorly explored despite the fact that Vasiliev system contains these tensors on the same footing as the symmetric ones. See, however, \cite{Basile:2015jjd,Joung:2016naf} and references therein.}. Indeed, these extra light-cone vertices are crucial for the consistency of the HS theories in four-dimensional Minkowski space \cite{Metsaev:1991mt} (see also \cite{Conde:2016izb,Devchand:1996gv,Sleight:2016xqq,Ponomarev:2016lrm,Ponomarev:2016cwi}).
The absence of corresponding covariant couplings in $d\geq 4$ is known as Aragone-Deser problem \cite{Aragone:1979hx} which is resolved in constant non-zero curvature $(A)dS$ spacetimes by the Fradkin-Vasiliev mechanism \cite{Fradkin:1987ks} (see \cite{Boulanger:2008tg,Joung:2013nma} for related discussion).

The covariant classification of cubic vertices in \cite{Manvelyan:2010jr} not only completed the light-cone vertices of Metsaev \cite{Metsaev:2005ar} to off-shell ones for Fronsdal fields but also defined a scheme of field redefinitions in a given cubic action to bring it to the form containing not more than $s_1+s_2+s_3$ derivatives. This form does not contain any contraction between derivatives and is uniquely defined. We refer to it as a vertex in Metsaev basis. This was implemented later in \cite{Boulanger:2015ova,Didenko:2015cwv} for translating the quadratic order of the Vasiliev equations in $(A)dS_4$, corresponding to cubic action, to the Metsaev basis in metric formulation, that is, $AdS$ extensions of Minkowski vertices for each number of derivatives $\Delta = s_1+s_2+s_3-2n$ for $n=0,1,\dots,min\{s_1,s_2,s_3\}$.

Attempts for going beyond cubic order
\cite{MM,Sagnotti:2010at,Dempster:2012vw,Bengtsson:2016hss,Taronna:2017wbx,Roiban:2017iqg} have met difficulties in the framework of local field theory. An interesting suggestion for a possibility of a non-local theory with conformal symmetry has been made in \cite{Roiban:2017iqg} which calls for further studies.

Another interesting recent development is the progress in the holographic reconstruction \cite{Bekaert:2014cea,Sleight:2016dba,Ponomarev:2017qab} of type A HS theory in $AdS_{d+1}$. Together with the aforementioned attempts of construction of a quartic order action via the Noether procedure, these results brought to the forefront of HS research the puzzle of locality which, to our best knowledge, was first posed sharply for three dimensional systems in \cite{Prokushkin:1998bq}. 
One may hope that the key to the solution of this puzzle can be found more easily in the three-dimensional case by applying recently obtained knowledge of the metric-like theory.
Unfortunately, most of the aforementioned advances in higher-dimensional HS gravities are not directly applicable to three dimensional models. This is due to the heavy use of Metsaev basis of cubic interactions in higher dimensions that does not apply to $d=3$. In order to make use of new results in metric-like HS gravity also for the three-dimensional models, one first needs to address the gap in the classification of cubic vertices. 
 In this paper, we continue the study aimed at filling this gap initiated in \cite{Mkrtchyan:2017ixk} where parity-even cubic vertices of massless bosons were classified. We complete the three-dimensional classification of cubic interactions deriving parity-odd vertices for massless bosonic fields as well as their couplings to Chern-Simons fields. We also elaborate on the analogous classification in two dimensions in the Appendix.

Despite all the successes of the three-dimensional HS gravities (see \cite{Prokushkin:1998bq,Banados:2016nkb,Campoleoni:2017xyl,Joung:2014qya,Iazeolla:2015tca} and references therein), there is no action formulation\footnote{See, however, \cite{Arias:2016ajh} and references therein for non-standard actions.} for the only known example of higher spin theory with propagating degrees of freedom in three dimensions i.e. Prokushkin-Vasiliev theory \cite{Prokushkin:1998bq}. This theory contains scalar degrees of freedom interacting with higher spin gauge fields which do not carry propagating degrees of freedom in the bulk.
The Chern-Simons formulation of HS gravity in three dimensions does not answer the question whether a Lagrangian for the Prokushkin-Vasiliev theory exists. This question may be tackled in the metric-like formulation where scalar and gauge fields can be put into interaction in a straightforward manner. This approach is much less explored in three dimensions though, with the exception of a few works on higher spins in the Fronsdal formulation \cite{Campoleoni:2012hp,Fredenhagen:2014oua,Campoleoni:2014tfa}.

S-matrix methods do not apply to three dimensions where massless particles of spin $s\geq 2$ do not propagate.
For the same reason, Metsaev's light-cone classification \cite{Metsaev:2005ar} does not work in three dimensions. The part of the cubic vertices that contains no divergences and traces i.e. traceless-transverse (TT) vertices are non-trivial though, as shown in \cite{Mkrtchyan:2017ixk}, and can serve as the basis for the classification of cubic interactions of massless fields in three dimensions.


The main difference between dimensions $d\geq 5$ and $d\leq 4$ for the cubic interactions of massless symmetric HS fields is the existence of dimension-dependent identities (Schouten identities) that are available in the latter case.
Due to these identities, the classification of cubic vertices in three dimensions becomes a completely independent problem which overlaps with the generic dimensional classification only for some vertices involving lower spin fields.
The classification of $d=3$ vertices was initiated in \cite{Mkrtchyan:2017ixk} where the parity-even vertices for interactions of massless bosons were derived. In this work, we complete the classification adding to it the parity-odd vertices of massless bosons as well as their interactions with Chern-Simons vector fields.

The paper is organized as follows: In Section \ref{FT} we review metric-like formulation of free HS fields. In Section \ref{Cubic Review} we review the construction of cubic vertices in higher dimensions, and the parity-even vertices in three dimensions. In Section \ref{PO} we derive full list of parity-odd cubic vertices of massless bosons in three dimensions and establish interesting relation between parity-odd and parity-even vertices. In Section \ref{Chern-Simons} we study interactions of massless fields with Chern-Simons vector fields. We conclude by summary of results and discussion in Section \ref{Discussion}. Appendices provide curious observations related to the parity-even vertices and classification of Fronsdal cubic vertices in two dimensions.

\section{Review: Free Theory}\label{FT}
In this paper we study interactions of massless fields of any spin as deformations of the free theory. To this end, we first set the stage by describing the free theory.
In order to streamline the notation, we will contract spacetime indices $\mu,\nu,\dots$ with commuting auxiliary variables $a^\mu$. In this language, the rank $s$ symmetric tensor field is given by:
\begin{align}
\phi^{\sst s}(a) = \frac{1}{s!} \, \phi_{\mu_1 \dots \mu_s} \, a^{\mu_1} \dots a^{\mu_s} \,.
\end{align}
In order to describe a free particle with spin $s$ in a covariant manner, one has to impose on the rank $s$ symmetric Lorenz tensor field the so-called Fierz equations \cite{Fierz}:
\begin{subequations}
\begin{align}
(\Box + m^2)\phi^{\sst s}(a)= \frac{1}{s!} \,(\Box + m^2)\,\phi_{\mu_1 \dots \mu_s} \, a^{\mu_1} \dots a^{\mu_s} =0\,,\label{Fierz1}\\
(\partial_x\cdot \partial_a)\phi^{\sst s}(a)=\frac1{(s-1)!}\,\partial^{\nu}  \phi_{\nu \mu_2 \dots \mu_s} \, a^{\mu_2} \dots a^{\mu_s} =0\,,\label{Fierz2}\\
\partial_a^2\phi^{\sst s}(a)=\frac{1}{(s-2)!} \, \phi^{\nu}{}_{\nu\mu_3 \dots \mu_s} \, a^{\mu_3} \dots a^{\mu_s}=0\,.\label{Fierz3}
\end{align}
\end{subequations}
For the massless fields $(m^2=0)$, one has to require also an extra equivalence between fields, differing by a gradient shift with traceless and transverse parameter $\e^{\sst s-1}(x;a)$:
\begin{align}
\d\phi^{\sst s}(a)=(a\cdot \partial_x)\e^{\sst s-1}(a)\,,\quad (\partial_x\cdot \partial_a)\e^{\sst s-1}(a)=0\,,\quad
\partial_a^2\e^{\sst s-1}(a)=0\,.
\end{align}
It has been a challenge to find a Lagrangian, even for the free Fierz equations. The natural expectations based on experience with lower spins is to have a single equation of motion for the rank $s$ tensor field, which has all the three Fierz equations as its consequences and also gauge symmetry of action in the massless case.

For the massless case, the most conventional description is due to Fronsdal \cite{Fronsdal:1978rb}.
The equation of motion is given by Fronsdal tensor:
\begin{align}
\mathcal{F}^{\sst s}(a) \equiv \left[ \Box - (a \cdot \partial_x) D \right] \, \phi^{\sst s}(a)=0 \,,
\end{align}
with the de Donder operator $D(a) = (\partial_x \cdot \partial_a) - \frac12 \, (a \cdot \partial_x) \partial^2_a$. 
The Fronsdal tensor  $\mathcal{F}$ is invariant with respect to gauge transformations:
\begin{align}
\delta \phi^{\sst s}(a) = (a \cdot \partial_x) \epsilon^{\sst s-1}(a) && \text{with} && \partial_a^2 \, \epsilon^{\sst s-1}(a) = 0 \,.  
\end{align}
The Fronsdal field $\phi^{\sst s}(a)$ is doubly traceless: 
\begin{align}
(\partial_a^2)^2 \, \phi^{\sst s}(a) =0 \,.
\end{align}
The action is given by:
\begin{align}
\label{eq:fronsdalAction}
S^{(s)} = \frac12 \int \text{d}^d x \; \phi^{\sst s}(a) (\overset{\leftarrow}{\partial_a} \cdot \overset{\rightarrow}{\partial_a} )^s \mathcal{G}^{\sst s}(a) \,,
\end{align}
with Lagrangian equations of motion:
\begin{align}
\mathcal{G}^{\sst s}(a) = \mathcal{F}^{\sst s}(a) - \frac14 \, a^2 \partial_a^2 \, \mathcal{F}^{\sst s}(a)=0 \,.
\end{align}
Using double-tracelessness of the Fronsdal field, one can easily show that the equations of motion $\mathcal{G}=0$ are equivalent to the Fronsdal equations $\mathcal{F}=0$.
At the linearised level, the Fronsdal equations imply the Fierz equations.

An alternative to the Fronsdal action is given by the Maxwell-like formulation \cite{Campoleoni:2012th} of HS dynamics.
The traceless-transverse (TT) parts of vertices in both formulations are equivalent though \cite{Francia:2016weg} and we will therefore not distinguish them in this work since we restrict ourselves to TT vertices only following \cite{Mkrtchyan:2017ixk}. The TT vertices studied here can be completed to off-shell vertices for both Fronsdal and Maxwell-like HS fields\footnote{It is an empirical observation that the TT vertices can be completed to off-shell ones, based on the known examples of both Fronsdal \cite{Manvelyan:2010jr} and Maxwell-like \cite{Francia:2016weg} vertices in $d\geq 4$. We do not have a proof that it will work in 3d straightforwardly. One interesting check would be to see if the ``Grassmann miracle'' of \cite{Manvelyan:2010je} (which allows to immediately derive the off-shell vertices from TT ones) works in this case. In three dimensions the off-shell vertex computations are technically involved, though, and we do not attempt them here.}. One can regard the results of this work as the classification of deformations of the Fierz system of equations \cite{Fierz} for massless HS fields in $d=3$. There is an important difference between Fronsdal and Maxwell-like descriptions relevant to this work which we do not elaborate on here, though. While Fronsdal fields do not carry propagating degrees of freedom in three dimensions, the reducible Maxwell-like fields do carry a propagating massless scalar (vector) degree of freedom for even (odd) spin. As a consequence, non-linear theories of Maxwell-like fields, if any, cannot be given by Chern-Simons actions in striking difference with many known models for Fronsdal fields. The classification that we carry out here can be implemented for building models with both Fronsdal and Maxwell-like field content. 

\section{Review: Cubic Vertices}\label{Cubic Review}
We will assume that there exists a gauge invariant non-linear action $S$ that can be expanded in power of fields with a small expansion parameter $g$ as follows
\begin{align}
S = S^{(2)} + g \, S^{(3)} + g^2 \, S^{(4)} + \dots \,,\label{FA}
\end{align}
where $S^{(2)} = S^{(s_1)} + S^{(s_2)} + S^{(s_3)}$ with $S^{(s_i)}$ denoting the Fronsdal action for the spin $s_i$ field \eqref{eq:fronsdalAction}.  
Gauge invariance of the action implies
\begin{align*}
\delta S = (\delta^{(0)} + g \delta^{(1)} + \dots ) (S^{(2)} + g S^{(3)} + \dots) && \rightarrow && \delta^{(0)} S^{(3)} + \delta^{(1)} S^{(2)} = 0 \,.
\end{align*}
Using the fact that $\delta^{(1)} S^{(2)} = \delta^{(1)} \phi \; \mathcal{G}$, it follows that\footnote{Note that our notation is somewhat schematic. The variation is to be understood as a sum of three terms, i.e. $\delta^{(1)} \phi \;\mathcal{G}= \delta^{(1)} \phi^{(s_1)} \, \mathcal{G}^{(s_1)}+ \delta^{(1)} \phi^{(s_2)} \, \mathcal{G}^{(s_2)}+ \delta^{(1)} \phi^{(s_3)} \, \mathcal{G}^{(s_3)}$.}
\begin{align}
\delta^{(0)} S^{(3)} \approx 0 \,,
\end{align}
where $\approx$ denotes equality upon imposing free equations of motion $\mathcal{G}=0$. Note also that any two actions $S$ and $S'$ related by a field redefinition $\phi \to \phi + g \, f(\phi,\phi)$, obey
\begin{align}
S^{(3)} \approx S'^{(3)} \,.
\end{align}
This ambiguity in field redefinition at the cubic order will be fixed by restricting the possibility of derivative contractions in the cubic vertex (as reviewed for example in \cite{Conde:2016izb,Mkrtchyan:2017ixk}).

One can now make the following ansatz for the cubic action
\begin{align}
S^{(3)} = \int \text{d}^dx \, \mathcal{V} \; \phi^{\sst s_1}(a_1,x_1) \, \phi^{\sst s_2}(a_2,x_2) \, \phi^{\sst s_3}(a_3,x_3) \; \left( \prod^3_{i=1} \delta(x-x_i) d^3x_i \right) \,,
\end{align}
where the differential operator $\mathcal{V}=\mathcal{V}(\partial_{x_1},\partial_{a_1},\partial_{x_2},\partial_{a_2},\partial_{x_3},\partial_{a_3})$ is to be determined. Since $\mathcal{V}$ is a scalar operator, it is built of contractions of the derivatives $\partial_{a_i}$ and $\partial_{x_i}$. It can be shown that, up to total derivatives and upon fixing the freedom in field redefinitions \footnote{The field redefinition freedom is fixed following \cite{Manvelyan:2010wp,Manvelyan:2010jr}, that is, by removing all terms with derivatives contracted with each other. Strictly speaking, one can exclude the derivative contractions by field redefinitions only in the terms that do not contain divergences. That turns out to be already sufficient for fixing the field redefinition freedom (see, e.g., \cite{Manvelyan:2010je}).}, all contractions can be written in terms of
\begin{subequations}
\begin{align}
\partial_{a_i}\cdot\partial_{x_i} &\equiv \text{Div}_i \,,\\
\partial_{a_i}\cdot\partial_{x_{i+1}} &\equiv y_i \,,\\
\partial_{a_i}\cdot\partial_{a_{i+1}} &\equiv z_{i+2} \,,\\
\partial_{a_i}\cdot\partial_{a_{i}} &\equiv T_i \,.
\end{align}
\end{subequations}
Let us furthermore restrict to interaction terms which do not involve traces $T_i$ or divergences Div$_i$. In this case, one obviously has
\begin{align}
\mathcal{V} = \mathcal{V}(y_i,z_i) \,.
\end{align}
The gauge variation of the ansatz for the cubic action is then given by
\begin{align*}
\delta^{(0)} S^{(3)} = \int \text{d}^d x \prod_{i=1}^{3} \text{d}x_i \delta(x-x_i) \; \mathcal{V} \; \sum_{j=1}^3 a_j \cdot \partial_{x_j} \epsilon(a_j,x_j) \; \phi(a_{j+1},x_{j+1}) \; \phi(a_{j-1},x_{j-1}) \,,
\end{align*}
where here and in the following we assume indices $i,j,\dots$ to be cyclic in $(1,2,3)$, for example $y_{i} \equiv y_{i+3}$.
Using the commutators
\begin{align}
[z_{i+1}, a_i \cdot \partial_i] \circeq y_{i+2}\,, && [z_{i+2}, a_i \cdot \partial_i] \circeq -y_{i+1} \,,
\end{align}
where $\circeq$ denotes equality up to equations of motion, total derivatives, traces and divergences. Similarly, it can be shown that all other commutators vanish up to these terms. After dropping total derivatives with respect to $\partial_{x_i}$, it then follows that 
\begin{align*}
\delta^{(0)} S^{(3)} = \int \text{d}^d x \prod_{i=1}^{3} \text{d}x_i \delta(x-x_i) \; \sum_j (y_{j-1} & \partial_{z_{j+1}}- y_{j+1} \partial_{z_{j-1}}) \; \mathcal{V}  \\ & \times \epsilon(a_j,x_j) \; \phi(a_{j+1},x_{j+1}) \; \phi(a_{j-1},x_{j-1})
\end{align*}
It then immediately follows that gauge invariant vertices solve the equations
\be
D_i \mathcal{V}\equiv (y_{i-1}\partial_{z_{i+1}}-y_{i+1}\partial_{z_{i-1}})\mathcal{V}=0\,,\quad i=1,2,3\,,\label{Di}
\ee
and are given by
\begin{align}
\mathcal{V}=\mathcal{V}(y_i,G)\qquad\qquad \text{with} \quad G = \sum_{i=1}^3 y_i \cdot z_i \,.
\end{align}
In generic spacetime dimension, these solutions span the entire space of possible cubic vertices. However at fixed dimension $d\le 4$, most of these vertices are vanishing \cite{Conde:2016izb}, while there may be more solutions due to Schouten identities as demonstrated in \cite{Mkrtchyan:2017ixk}. We will briefly review the main results of the latter work on parity-even vertices in $d=3$ here.

\subsection{Parity-even Vertices in $d=3$}

The derivation of Lorentz covariant cubic vertices described above has to be supplemented with Schouten identities that are relevant for $d\le 4$. The case of four dimensions can be found in \cite{Conde:2016izb} while in \cite{Mkrtchyan:2017ixk} the $d=3$ parity even vertices were classified. The Schouten identities can be systematically derived by ``over-antisymmetrisation'' of Lorentz indices and there are even Mathematica packages doing so \cite{Nutma:2013zea}.
The elementary three-dimensional Schouten identities for parity-even TT cubic vertices are given as (grouped in two-, three- and four-derivative identities, no summation over repeating indices assumed):
\begin{subequations}\label{SI}
\begin{align}
(G - y_i z_i)^2 = 0 \,, && y_i z_i G - y_{i-1} z_{i-1} y_{i+1} z_{i+1} = 0 \,, \label{eq:twoDerivDDI}\\
y_i y_{i\pm 1}(G - y_i z_i) = 0 \,, \label{eq:threeDerivDDI}\\
y_i^2 y^2_{i+1} = 0 \,, && y_i^2 y_{i+1} y_{i-1}=0 \label{eq:fourDerivDDI}\,.
\end{align} 
\end{subequations}
These identities will be supplemented with parity-odd ones in the next section and are needed for the derivation of parity-odd vertices.

Due to identities \eqref{SI}, the classification of parity-even cubic vertices in 3d is different from that of $d\geq 4$. In particular, these identities allow for existence of two-derivative and three-derivative TT vertices given by \cite{Mkrtchyan:2017ixk}:
\begin{subequations}\label{PE}
\begin{framed}
\begin{align}
\mathcal{V}_{s_1,s_2,s_3} &=[(s_1-1) y_1 z_1+(s_2-1) y_2 z_2+(s_3-1) y_3 z_3] G z_1^{n_1} z_2^{n_2} z_3^{n_3}\,,\label{2vertex}\\
n_i&=\tfrac12(s_{i-1}+s_{i+1}-s_i)-1\geq 0\,,\nonumber\\
\mathcal{V}_{s_1,s_2,s_3} &= y_1\, y_2\, y_3\, z_1^{n_1}\, z_2^{n_2}\, z_3^{n_3}\,,\qquad
n_i=\tfrac12(s_{i-1}+s_{i+1}-s_i-1)\geq 0\,.\label{3vertex}
\end{align}
\end{framed}
\end{subequations}
The expressions \eqref{2vertex} and \eqref{3vertex} describe unique cubic vertices for even and odd sum of spins respectively. Note that \eqref{2vertex} involves minimal coupling to gravity discussed for particular cases earlier in \cite{Aragone:1983sz,Gwak:2015vfb,Campoleoni:2012hp}. These vertices exist only if the spin values satisfy triangle inequalities $s_i< s_{i+1}+s_{i-1}$.

\section{Parity-odd Vertices for Massless Bosons}\label{PO}

In order to construct parity-odd vertices of massless fields in three dimensions, one needs to add to the building blocks of the parity-even vertices, i.e. $y_i$ and $z_i$, all elementary scalar contraction operators that involve the invariant tensor $\epsilon_{\mu\nu\lambda}$ of the Lorentz algebra. These are:
\begin{align}
U=\epsilon^{\mu\nu\lambda}\partial^{a_1}_{\mu}\partial^{a_2}_{\nu}\partial^{a_3}_{\lambda}\,,\quad V_{ij}=\epsilon^{\mu\nu\lambda}\partial^{a_{i+1}}_{\mu}\partial^{a_{i-1}}_{\nu}\partial^{x_j}_{\lambda}\,,\quad W_i=\epsilon^{\mu\nu\lambda}\partial^{a_i}_{\mu}\partial^{x_{i+1}}_{\nu}\partial^{x_{i-1}}_{\lambda}\,,
\end{align}
where the $V$'s satisfy (discarding total derivative terms)
\begin{align}
\sum_{j}V_{ij}=0\,,
\end{align}
while the $W$'s are a choice of basis for nine different structures with two derivatives related to each other up to total derivatives.
Therefore the independent set of parity-odd variables is spanned by ten scalar operators $U, V_{ij} (i\neq j), W_i$.
It is straightforward to check that:
\begin{align}
[U, a_i\cdot\partial_j]=V_{ij}\,,\quad [V_{i i\pm 1}, a_i\cdot \partial_j]=0\,,\quad
[V_{i i\pm 1},a_{j}\cdot\partial_{i\pm 1}]=0\,,\\
[V_{i i\pm 1},a_{i\mp 1}\cdot\partial_{i\mp 1}]=-W_{i\pm 1}\,,\quad
[V_{i i\pm 1}, a_{i\pm 1}\cdot \partial_{i\mp 1}]=\pm W_{i\mp 1} \,,\\
[V_{i i\pm 1}, a_{i\pm 1}\cdot \partial_{i}]=-W_{i\mp 1} \,,\quad
[V_{i i\pm 1}, a_{i\mp 1}\cdot \partial_{i}]=W_{i\pm 1} \,,\\
 [W_i, a_j\cdot\partial_k]=0\,,
\end{align}
up to total derivatives.

The operator $D_i$ \eqref{Di} takes the following form for parity-odd vertices:
\begin{align}
D_i = y_{i-1} \partial_{z_{i+1}} - y_{i+1} \partial_{z_{i-1}} - W_{i-1} \partial_{V_{i+1 i-1}} - W_{i+1} \partial_{V_{i-1 i+1}}\nonumber\\ - V_{i i-1} \partial_U - V_{i i +1} \partial_U \,.
\end{align}
The elementary parity-odd Schouten identities are given by (with arbitrary $i,j,k$ and no summation over repeating indices assumed):
\begin{subequations}\label{POSI}
\begin{align}
V_{i-1 i} z_{i+1} + V_{i+1 i} z_{i-1} =0\,,\quad
U y_i + V_{i+1 i-1} z_{i-1} - (V_{i-1 i} + V_{i-1 i+1})z_{i+1} =0\,,\label{1dSI}\\
W_{i+1} z_{i+1} - W_{i} z_{i} = V_{j i-1} y_{j}=V_{k i-1} y_{k}\,,\quad
W_{i} z_{i+1} = V_{i+1 i} y_{i}\,,\quad
W_{i} z_{i-1} = - V_{i-1 i} y_{i}\,,\label{2dSI} \\
W_{i} y_{i\pm1} = 0\,,\label{3dSI}
\end{align}
\end{subequations}
where identities are grouped into one, two and three derivative ones. From these we derive other useful identities:
\begin{align}
V_{i i\pm 1}(y_i z_i+y_{i\mp 1} z_{i\mp 1})= 0\,,\label{Vyz}\\
W_i\, y_i\, z_i=V_{i i+1}\,y_i^2=-V_{i i-1}\,y_i^2\,,\quad V_{ij}y_j y_{j\pm1}= 0\,,\quad W_i\,y_i\,z_i^2=- U\,y_1\,y_2\,y_3\,,\label{Wyzz}\\
V_{i i\pm1}y_i^2 y_{i\mp 1}= 0\,,\quad V_{ij}\,y_1\,y_2\,y_3=0\,,\quad W_i y_i^2 z_i^2= 0\,.\label{Vyyy}
\end{align}
A consequence of the Schouten identities is that all parity-odd terms with more than one derivative can be written in terms of $W_i,\,y_i,\,z_i$ operators only as long as the spins satisfy triangle inequalities $s_i< s_{i+1}+s_{i-1}$. This property will be useful in the following. The terms that cannot be written only in terms of the variables mentioned above are of the following form:
\begin{align}
V_{i i\pm 1}y_{i\mp 1}^n z_i^m z_{i\pm 1}^p\neq \sum_k \mathcal{O}(W_k)\,.\label{NW}
\end{align}
We assume without loss of generality $s_1\geq s_2\geq s_3$. Then, all the terms of the type \eqref{NW} are given as:
\begin{align}
V_{23}\,y_1^{s_1-s_2-s_3}z_2^{s_3-1}z_3^{s_2}\,,
\qquad
V_{32}\,y_1^{s_1-s_2-s_3}z_2^{s_3}z_3^{s_2-1}\,,\label{nW}
\end{align}
and exist only for $s_1 \geq s_2+s_3$.
We also note that any expression written in terms of $W_i$ and non-vanishing up to identities \eqref{3dSI} and \eqref{Vyyy} cannot be converted to $V_{ij}$ expressions that vanish. This is due to the fact that
terms involving $W_k$ with different $k$ give rise to $V_{ij}$ with different $j$ that cannot sum up to zero through identities \eqref{1dSI} and \eqref{2dSI}. This simple technical observation suggests that working solely with $W_k$-s wherever possible will not miss any information about terms that may conspire to sum up to zero. Since $W_i$-s are also commuting with all gauge variations, it makes them the preferred choice of variables in expressions with more than one derivative that we will study in the following.

We now proceed with the derivation of parity-odd cubic vertices for massless bosons in three dimensions. We will need to discuss separately different cases and simplify each ansatz maximally, to save virtual trees that get cut in order to supply us with Mathematica notebooks.

\subsection{Vertices with Scalars}

The simplest example is a vertex with two scalar fields involved: $(s,0,0)$.
In this case the only candidate vertex operator is:
\begin{align}
\boxed{\mathcal{V}_{s,0,0}^{PO}=W_1\,y_1^{s-1}\,,}\label{POs00}
\end{align}
and defines a gauge invariant vertex of current interaction type:
\begin{align}
\mathcal{L}_{s,0,0}=h^{\mu_1\dots\mu_s}\tilde{J}_{\mu_1\dots\mu_s}\,,
\end{align}
where the current
\begin{align}
\tilde{J}_{\mu_1\dots\mu_s}=\epsilon_{\nu\rho(\mu_1}\partial^{\nu}J^{\rho}{}_{\mu_2\dots\mu_s)}\,,
\end{align}
is roughly the curl of the parity-even conserved current $J_{\mu_1\dots\mu_s}$ of spin $s$.
Next we look at the possible vertices with $s_1\geq s_2\geq 1$ and $s_3=0$.
The general ansatz for $(s_1,s_2,0)$ vertex can be written as:
\begin{align}
\mathcal{V}_{s_1,s_2,0}=(\alpha V_{31} + \beta V_{32})y_1^{s_1-s_2}z_3^{s_2-1}\,.\label{POs1s20}
\end{align}
The variation of \eqref{POs1s20} with respect to the gauge symmetry of the spin $s_1$ field gives:
\begin{align}
D_1 \mathcal{V}_{s_1,s_2,0}=(s_2-1)(\alpha V_{31}+\beta V_{32})\,y_1^{s_1-s_2}\,y_2\,z_3^{s_2-2}-\beta w_2\,y_1^{s_1-s_2}\,z_3^{s_2-1}\,,
\end{align}
The variation of \eqref{POs1s20} with respect to the symmetry of the spin $s_2$ field gives:
\begin{align}
D_2 \mathcal{V}_{s_1,s_2,0}&=-\alpha W_1 y_1^{s_1-s_2}z_3^{s_2-1}+(s_2-1)(\alpha V_{31} + \beta V_{32})y_1^{s_1-s_2+1}z_3^{s_2-2}\nonumber\\
&=-s_2\,\alpha\, W_1 y_1^{s_1-s_2}z_3^{s_2-1}+(s_2-1)\, \beta\, V_{32}y_1^{s_1-s_2+1}z_3^{s_2-2}\,.\label{VarPOs1s20}
\end{align}
Vanishing of this variation is compatible with a non-zero vertex only for $s_2=1$, $\alpha=0$. Therefore, there is a unique vertex:
\begin{align}
\boxed{\mathcal{V}_{s,1,0}^{PO}=V_{32}\,y_1^{s-1}\,,}\label{POs10}
\end{align}
which is invariant with respect to the gauge transformation of the second (Maxwell) field ($D_2 \mathcal{V}_{s,1,0}^{PO}= 0$), while the gauge variation of the spin $s$ field,
\begin{align}
D_1 \mathcal{V}_{s,1,0}^{PO}= - W_2 y_1^{s-1}\,,
\end{align}
vanishes due to \eqref{3dSI} iff $s\geq 2$. Therefore, the vertex of type $(s,1,0)$ exists for any $s\geq 2$.

For $s_2\geq 2$, the variation \eqref{VarPOs1s20} can be rewritten as:
\begin{align}
D_2\mathcal{V}_{s_1,s_2,0}=(s_2\, \alpha\, V_{31} + (s_2-1)\, \beta\, V_{32})y_1^{s_1-s_2}z_3^{s_2-2}\,,
\end{align}
and allows for gauge invariance only for trivial solution $\alpha=\beta=0$.

Similar to the parity-even case, there are no couplings of the type $(s_1,s_2,0)$ with $s_1\geq s_2\geq 2$.
Thus we found all the vertices involving scalar fields.

\subsection{Vertices with Maxwell Fields}

From the Schouten identity \eqref{3dSI}, it follows immediately that there is a vertex of the type $(s,s,1)$ with two derivatives:
\begin{align}
\boxed{\mathcal{V}_{s,s,1}^{PO}=W_3\, z_3^s\,.}\label{PO1ss}
\end{align}
It is a parity-odd two-derivative coupling to spin one which requires charged spin-$s$ fields. For $s=1$, \eqref{PO1ss} reproduces the spin one vertex found by Anco in \cite{Anco:1995wt}.
There is another vertex of the type $(s,1,1)$ that may be guessed immediately:
\begin{align}
\boxed{\mathcal{V}_{s,1,1}^{PO}=W_1\,y_1^{s-1}\, z_1\,,}\label{POs11}
\end{align}
which involves $s+1$ derivatives. For $s=1$, \eqref{POs11} coincides with \eqref{PO1ss} up to relabelling of the fields. We will come back to this vertex shortly.

It remains to check other possibilities of interactions $s_1\geq s_2\geq s_3=1$ with Maxwell field.
It is straightforward to see that the number of derivatives cannot be less than $s_1-s_2$ simply because there are no candidate scalar expressions. The upper bound on derivatives is a bit more subtle to define. An obvious upper bound is $s_1+s_2$, since all vertex monomials with $s_1+s_2+2$ derivatives vanish due to \eqref{3dSI} for any $s_1,\,s_2$ and there are no candidate expressions with derivatives more than $s_1+s_2+2$.
Nevertheless, it can be easily shown that for $s_1\geq s_2>>1$ the upper bound is much lower than $s_1+s_2$ due to \eqref{eq:fourDerivDDI} and \eqref{3dSI}.
In fact, careful examination taking into account all Schouten identities shows that there are no non-trivial vertex candidates for the number of derivatives more than $s_1-s_2+2$.
Therefore we are left with two candidate values for number of derivatives in the vertex: $s_1-s_2$ and $s_1-s_2+2\,.$
We will consider these cases separately.

\paragraph*{$(s_1-s_2+2)-$ derivative vertex.}
With the help of some elementary algebra and making use of Schouten identities, a general ansatz for an $s_1-s_2+2$ derivative vertex can be written in the form:
\begin{align}
\mathcal{V}_{s_1,s_2,1}=[\gamma_1 W_1 z_1
+\gamma_2 W_2 z_2+\gamma_3 W_3 z_3]y_1^{s_1-s_2}z_3^{s_2-1}
\,,\label{POs1s21}
\end{align}
where $\gamma_i$ are arbitrary constants.

For simplicity, we discuss separately the cases of $s_2=1$, $s_1=s_2$ and  $s_1>s_2\geq 2$.

\begin{itemize}

\item
For $s_2=1$, we have a general ansatz
\begin{align}
\mathcal{V}_{s,1,1}=[\gamma_1 W_1 z_1
+\gamma_2 W_2 z_2+\gamma_3 W_3 z_3]y_1^{s-1}\,.
\end{align}

For $s=1$, we have:
\begin{align}
\boxed{\mathcal{V}_{1,1,1}^{PO}=\gamma_1 W_1z_1+\gamma_2 W_2z_2+\gamma_3 W_3z_3
\,,}\label{PO111}
\end{align}
with arbitrary $\gamma_i$. Each of the three terms in this expression is separately gauge invariant and defines a vertex of the type \eqref{PO1ss}. We have three inequivalent vertices, defined for any triple of Maxwell fields. As opposed to the Yang-Mills vertex, which is fully antisymmetric in all the three fields involved, the term $W_iz_i$ is antisymmetric only in two fields $A^{i\pm 1}_\mu$ and can even define a cubic vertex for only two distinct Maxwell fields (e.g. taking value in the two-generator Lie algebra of infinitesimal affine transformations of a real line).
This vertex has been studied in \cite{Anco:1995wt}. One can write it in explicit form:
\begin{align}
\mathcal{L}_{1,1,1}=f_{abc}\epsilon^{\mu\nu\lambda}A^a_\mu \tilde{F}^b_\nu\tilde{F}^c_\lambda\,,\quad \tilde{F}^a_\mu=\epsilon_{\mu\nu\rho}\partial^\nu A^{a \rho}\,,\quad f_{abc}=-f_{acb}\,.
\end{align}

For $s\geq 2$, we have:
\begin{align}
\mathcal{V}_{s,1,1}=\gamma_1 W_1 z_1
y_1^{s-1}\,.
\end{align}
This is the $(s,1,1)$ vertex \eqref{POs11} where, for odd $s$, non-trivial interaction requires charged Maxwell fields.

\item

For $s_1=s_2=s>1$, the general ansatz with $s_1-s_2+2=2$ derivatives is:
\begin{align}
\mathcal{V}_{s,s,1}=[\gamma_1 W_1 z_1
+\gamma_2 W_2 z_2+\gamma_3 W_3 z_3]z_3^{s-1}\,,
\end{align}
 and is gauge invariant iff $\gamma_1=\gamma_2=0$. Therefore, we end up with the unique possibility of the vertex \eqref{PO1ss}.
 
\item
For $s_1>s_2\geq 2$, the general ansatz \eqref{POs1s21} reduces to:
\begin{align}
\mathcal{V}_{s_1,s_2,1}=\gamma_1 W_1 z_1
y_1^{s_1-s_2}z_3^{s_2-1}
\,,
\end{align}
which is not gauge invariant under the variation of the second field with spin $s_2$.
\end{itemize}

Therefore we find that all the vertices for $s_1\geq s_2\geq s_3=1$ with $s_1-s_2+2$ derivatives are covered by \eqref{PO1ss}, \eqref{POs11} and \eqref{PO111}.

\paragraph*{$(s_1-s_2)-$ derivative vertex.}

A general ansatz with $s_1-s_2$ derivatives can be written for $s_1=s_2=s$ in the form (without derivatives since $s_1-s_2=0$):
\begin{align}
\mathcal{V}_{s,s,1}=U z_3^{s-1}\,.
\end{align}
It is elementary to check that this expression is not gauge invariant. We will assume in the following that $s_1>s_2$, in which case the general ansatz takes the form:
\begin{align}
\mathcal{V}_{s_1,s_2,1}=(\alpha V_{23} z_3+\beta V_{21} z_3+\gamma V_{32}z_2) y_1^{s_1-s_2-1} z_3^{s_2-1}\,,
\end{align}
and it is easy to check that the equation $D_1 \mathcal{V}_{s_1,s_2,1}=0$ have only vanishing solutions for the coefficients $\alpha, \beta, \gamma$ unless $s_2=1$, $\beta=0,\, \gamma=-\alpha$.
The only candidate expression
$\mathcal{V}_{s,1,1}=(V_{23} z_3-V_{32}z_2)y_1^{s-2}\,$
is however not invariant with respect to the gauge transformations of Maxwell fields. 

We conclude that all the parity-odd vertices with Maxwell fields are given by \eqref{PO1ss}, \eqref{POs11} and \eqref{PO111}.

\subsection{Gravitational Interactions}

Making use of the Schouten identity \eqref{3dSI}, one can easily show that there is a three-derivative parity-odd  $(s,s,2)$ coupling to massless spin two:
\begin{align}
\boxed{\mathcal{V}_{s,s,2}^{PO}=W_3\, y_3\, z_3^s\,.}\label{PO2ss}
\end{align}
This vertex is symmetric with respect to the exchange of spin $s$ fields and therefore does not require charged fields.
For $s=2$, the expression \eqref{PO2ss} reproduces the vertex found by Boulanger and Gualtieri in \cite{Boulanger:2000ni}.

Any $(s,s,2)$ type of parity-odd vertex requires an odd number of derivatives.
We will see in the following that there are no parity-odd vertices with one derivative or with more than four derivatives. Therefore, \eqref{PO2ss} is the unique parity-odd coupling to gravity for given spin $s$.

We conclude that the parity-odd minimal coupling to gravity is given by a three-derivative vertex. Let us recall that the parity-even gravitational coupling has two derivatives. This is in contrast to spin one (Maxwell) minimal couplings, where the parity-odd coupling \eqref{PO1ss} has two derivatives while the parity-even coupling has three derivatives \cite{Mkrtchyan:2017ixk}.

It remains to see what the other options of $s_1> s_2\geq s_3=2$ couplings are.
These vertices will be classified in the following where we will consider the more general case of couplings between fields with arbitrary spin values.

\subsection{General Case}

It is elementary to verify, by making use of Schouten identity \eqref{3dSI}, that the expression $W_1 y_1^n z_1^m$ is gauge invariant for any $n,m$ and therefore forms a vertex. These type of vertices are all exhausted by \eqref{POs00}, \eqref{PO1ss}, \eqref{POs11} and \eqref{PO2ss}. There are no vertices of the aforementioned type with $n,m\geq 2$ due to \eqref{Vyyy}.

After the examples with low spins, we now start studying more general cases of cubic interactions.

\subsubsection{Couplings without Derivatives}

It is straightforward to show that there are no vertices without derivatives. In order to do so, one just needs to gauge variate the most general ansatz, 
\begin{align}
\mathcal{V}_{s_1, s_2, s_3}=U\,z_1^{n_1}\,z_2^{n_2}\,z_3^{n_3}\,,\quad s_i=n_{i-1}+n_{i+1}+1\,,
\end{align}
and compare to the linear combination of one-derivative Schouten identities with arbitrary coefficients. One will thus verify that there is no such linear combination and therefore no vertex without derivatives. 

\subsubsection{One-derivative Vertices}

It is straightforward to show that any one-derivative parity-odd vertex with three massless fields of spins $s_1\geq s_2\geq s_3\geq 2$ could be written in the following form:
\begin{align}
\mathcal{V}_{s_1,s_2,s_3}=\sum_{i=1}^3 (\alpha_i V_{i i+1}+\beta_i V_{i i-1})z_{i+1} z_{i-1} z_1^{n_1} z_2^{n_2} z_3^{n_3}\,,
\end{align}
which can be further simplified in case if the spins satisfy triangle inequalities $s_1<s_2+s_3$ to
\begin{align}
\mathcal{V}_{s_1,s_2,s_3}=(\alpha_1 V_{12} z_2 z_3+\alpha_2 V_{23} z_3 z_1+\alpha_3 V_{31} z_1 z_2) z_1^{n_1} z_2^{n_2} z_3^{n_3}\,,
\end{align}
and, if they saturate triangle inequality $s_1=s_2+s_3$, to
\begin{align}
\mathcal{V}_{s_1,s_2,s_3}= (\alpha V_{23}z_3+\beta V_{32}z_2) z_2^{n_2} z_3^{n_3}\,.
\end{align}
In both cases, even though there are solutions for each of the equations $D_i \mathcal{V}_{s_1,s_2,s_3}=0$, there is no non-zero intersection between these solutions\footnote{This observation may be useful in classification of couplings with massless {\it and massive} fields (since massive fields are not constrained by gauge invariance), which is out of the scope of this work.}.
It is also easy to verify that there are no candidate expressions for one-derivative vertices if $s_1>s_2+s_3$. 

We conclude that there are no parity-odd vertices with one derivative for any spins $s_1\geq s_2\geq s_3\geq 0$. 

\subsubsection{Two-derivative Vertices}

Now we turn to studying two derivative parity-odd interactions. This corresponds to odd values of the sum $s_1+s_2+s_3$ and therefore the triangle inequality cannot be saturated: $s_1\neq s_2+s_3$.
We discuss separately the case when the spins satisfy triangle inequalities and when they do not. 

\subparagraph{Triangle inequalities are satisfied.}
Taking into account that for $s_1\geq s_2\geq s_3\geq 1$ and $s_1<s_2+s_3$, any vertex monomial with two derivatives can be brought to the form where the only parity-odd operators are $W_i$. We end up with a simple ansatz:
\begin{align}
\mathcal{V}_{s_1,s_2,s_3}=[\alpha W_1z_1+\beta W_2z_2+\gamma W_3z_3]\,z_1^{n_1}z_2^{n_2}z_3^{n_3}\,, \quad n_1\leq n_2\leq n_3\,.
\end{align}
Making use of \eqref{3dSI} and \eqref{Wyzz}, one can show that
the gauge invariance conditions,
\begin{subequations}
\begin{align}
D_1\mathcal{V}_{s_1,s_2,s_3}=[-\b\, n_3\,W_2\,y_2\,z_2^2+\g\, n_2\, W_3\,y_3\,z_3^2]\,z_1^{n_1}z_2^{n_2-1}z_3^{n_3-1}\nonumber\\
=(\b\,n_3-\g\,n_2)\,U\,y_1\,y_2\,y_3\,z_1^{n_1}z_2^{n_2-1}z_3^{n_3-1}=0\,,\\
D_2\mathcal{V}_{s_1,s_2,s_3}=[-\g\,n_1\,W_3\,y_3\,z_3^2+\a\, n_3\, W_1\,y_1\,z_1^2]\,z_1^{n_1-1}z_2^{n_2}z_3^{n_3-1}\nonumber\\
=(\g\,n_1-\a\,n_3)\,U\,y_1\,y_2\,y_3\,z_1^{n_1-1}z_2^{n_2}z_3^{n_3-1}=0\,,\\
D_3\mathcal{V}_{s_1,s_2,s_3}=[-\a\,n_2\,W_1\,y_1\,z_1^2+\b\, n_1\, W_2\,y_2\,z_2^2]\,z_1^{n_1-1}z_2^{n_2-1}z_3^{n_3}\nonumber\\
=(\a\,n_2-\b\,n_1)\,U\,y_1\,y_2\,y_3\,z_1^{n_1-1}z_2^{n_2-1}z_3^{n_3}=0\,,
\end{align}
\end{subequations} 
imply:
\begin{align}
\b\,n_3-\g\,n_2=\g\,n_1-\a\,n_3=\a\,n_2-\b\,n_1=0\,.\label{n1n2n3abg}
\end{align}
The solution to these equations fixes the vertex uniquely up to an overall constant\footnote{The vertex is not unique only when $n_1=n_2=n_3=0$. In this case, the equations \eqref{n1n2n3abg} are trivialised and there are no restrictions on $\a,\b,\g$. This case corresponds to the vertex given by the equation \eqref{PO111}.}:
\begin{align}
\boxed{\mathcal{V}_{s_1,s_2,s_3}^{PO}=[n_1\, W_1z_1+n_2\, W_2z_2+n_3\, W_3z_3]\,z_1^{n_1}z_2^{n_2}z_3^{n_3}\,, \quad s_i=n_{i+1}+n_{i-1}+1\,.}\label{2dPO}
\end{align}
This vertex exists for any triples of spins with odd sum satisfying strict triangle inequalities. For $s_1=s_2=s_3=3$, the expression \eqref{2dPO} reproduces the vertex found by Boulanger, Leclerq and Cnockaert in \cite{Boulanger:2005br}. To our best knowledge, the latter is the only example known in literature of parity-odd cubic vertices of HS fields in three dimensions.

This result is similar to parity-even case, where the two-derivative vertex \eqref{2vertex} exists for every triple of spins, with even sum, satisfying strict triangle inequalities. One important difference is that if there are two fields with the same spins in the vertex the parity-even vertex with two derivatives, \eqref{2vertex}, is symmetric with respect to exchange of these fields while parity-odd one, \eqref{2dPO}, is antisymmetric (we assume at least one of the spins is greater than one). We will come back to the relation of the parity-even and parity-odd vertices in the following. 

\subparagraph*{Triangle inequalities are violated.}

It is elementary to show that for $s_1> s_2+s_3+1$ there are no vertex monomials with two derivatives.
The only allowed case is $s_1=s_2+s_3+1$, with an ansatz involving expressions of the type \eqref{nW}:
\be
\mathcal{V}_{s_1,s_2,s_3}^{PO}=(\a V_{23} z_3+\b V_{23} z_2)y_1\,z_2^{s_3-1}\,z_3^{s_2-1}+\g W_1\,z_2^{s_3}\,z_3^{s_2}\,.
\ee
This expression is invariant with respect to the spin $s_1$ field's gauge variation, $D_1\mathcal{V}=0$, if $\a\, s_2+\b\, s_3=0$, and with respect to the second field's gauge variation, $D_2\mathcal{V}=0$, if $\a\, s_2=0=\b\, (s_2-1)\,,\,\, \g s_2=0$, while for the invariance with respect to the third field we get: $\a\,(s_3-1)=0=\b\,s_3\,,\,\, \g\, s_3=0$.
The only non-trivial solutions for this class of vertices are given by \eqref{POs00} with $s=1$ and \eqref{POs10} with $s=2$. 

We conclude that, similarly to parity-even vertices with two derivatives, there is only one parity-odd vertex \eqref{2dPO} with two derivatives for each triple of spins $s_1\geq s_2\geq s_3\geq 2$ satisfying triangle inequalities $s_1<s_2+s_3$ and with odd sum $s_1+s_2+s_3$.

\subsubsection{Three-derivative Vertices}

For vertices with three derivatives, we consider separately three cases depending on the values of spins. This corresponds to even sum of the spins.

\subparagraph*{Triangle inequalities are satisfied.}

In this case, the general ansatz for the vertex is given by (the overall arbitrary coefficient is dropped):
\ba
\boxed{\mathcal{V}_{s_1,s_2,s_3}^{PO}= -W_i\,y_i\,z_i^2\, z_1^{n_1}\,z_2^{n_2}\,z_3^{n_3}=U\,y_1\,y_2\,y_3\,z_1^{n_1}\,z_2^{n_2}\,z_3^{n_3}\,,\quad s_i=n_{i-1}+n_{i+1}+2\,.\label{PO3}}
\ea
Remarkably, this vertex is gauge invariant with respect to all three variations due to \eqref{eq:fourDerivDDI} and \eqref{Vyyy}.

This result is similar to the parity-even case \eqref{PE}, where every triple of spins with odd sum defined a unique vertex \eqref{3vertex} proportional to $y_1\,y_2\,y_3$ \footnote{Except for $(1,1,1)$ Yang-Mills fields, for which there were two vertices --- with one derivative and three derivatives.}. One notable difference is that in case if there are two fields with identical spin, due to the factor $U$, the vertex is symmetric with respect to permutations of these fields, as opposed to the vertex \eqref{3vertex}, which would be antisymmetric. For $s_1=s_2=s_3=2$ the vertex \eqref{PO3} reproduces the symmetric $d=3$ vertex of \cite{Boulanger:2000ni}.

\subparagraph*{Triangle inequalities are saturated.}

In this case, there are no non-trivial vertex monomials with three derivatives.

\subparagraph*{Triangle inequalities are violated.}

This case allows for non-trivial vertex ansatz iff $s_1=s_2+s_3+2$. The most general ansatz is given by:
\ba
\mathcal{V}_{s_1,s_2,s_3}^{PO}=(\a\, V_{23} z_3+\b\, V_{32}\, z_2)y_1^2\,z_2^{s_3-1}\,z_3^{s_2-1}+\g W_1\, y_1\,z_2^{s_3}\,z_3^{s_2}\,.
\ea
The analysis of this case is similar to the two-derivative one performed above. The only non-trivial solutions have been covered by \eqref{POs00} with $s=2$ and \eqref{POs10} with $s=3$.

It is a straightforward algebraic exercise to show that there are no non-trivial vertices with more than three derivatives for $s_1\geq s_2\geq s_3\geq 2$. This completes the classification of parity-odd cubic vertices of massless bosonic fields in three space-time dimensions.

\subsection{Relations between Parity-Odd and Parity-Even Vertices}

There is a remarkable universality in the formulas of the vertices \eqref{2vertex}, \eqref{2dPO}, \eqref{3vertex} and \eqref{PO3}. In order to show it, we first notice the following relation (as always in this work, we neglect trace terms):
\be
U^2=-2\, z_1\,z_2\,z_3\,.
\ee
Now, we can formally define the following operator 
\be
z_1^{1/2}\,z_2^{1/2}\,z_3^{1/2}=\frac{i}{\sqrt{2}}U\,.
\ee
Now, let us shift the integers $n_i$ in the \eqref{2dPO} by half: $n_i\to n_i+\frac12$.
Then the sum of the spins becomes even and the equation \eqref{2dPO} can be formally rewritten as:
\begin{align}
\mathcal{V}_{s_1,s_2,s_3}
=\frac{i}{\sqrt{2}}\Big[(n_1+\frac12)\, W_1z_1+(n_2+\frac12)\, W_2z_2+(n_3+\frac12)\, W_3z_3\Big]\,U\,z_1^{n_1}z_2^{n_2}z_3^{n_3}\nonumber\\
=\frac{i}{\sqrt{2}}[(s_1-1) y_1 z_1+(s_2-1) y_2 z_2+(s_3-1) y_3 z_3]\, G\, z_1^{n_1} z_2^{n_2} z_3^{n_3}\,,\label{2dPE}
\end{align}
where  $s_i=n_{i+1}+n_{i-1}+2\,$. Here we used the identities:
\be
W_i\,U=y_i(G-y_i\,z_i)\,,
\ee
and \eqref{eq:twoDerivDDI}.
The equation \eqref{2dPE} exactly reproduces the parity-even vertex \eqref{2vertex} up to an overall constant.
It is elementary to show, that the same relation holds between three-derivative parity-odd \eqref{PO3} and parity-even \eqref{3vertex} vertices.

Another curiosity related to parity-even vertices is discussed in Appendix \ref{A}.

This universality in formulas may have a deeper meaning in terms of certain dualities between fields that is yet to be uncovered. For example, it may be related to the Chern-Simons formulation where each HS field has two connections analogous to dreibein and spin connection of gravity. When switching to the Fronsdal formulation, the ``spin connection'' is solved in terms of the ``frame field'' and the solution contains one derivative and a Levi-Civita tensor. Replacing one ``frame field'' with a ``spin connection'' partner may result in switching the interactions between parity-odd and parity-even ones (it changes the parity and the number of derivatives by one). This is a speculation but can be checked by explicit computations. It is also tempting to speculate about the existence of a more fundamental formulation of any HS theory in terms of spinors, analogous to \cite{Prokushkin:1998bq,Vasiliev:1990en}, that treat the parity-even and parity-odd vertices on the same footing.
We leave more thorough investigations of this aspect to a future work.

\section{Vertices with Chern-Simons Vector Fields}\label{Chern-Simons}

So far we have been studying TT vertices of Fierz-type fields including the Maxwell field for $s=1$ that is given by the free Lagrangian:
\be
\mathcal{L}^0_{s=1}=-\frac14 F_{\m\n}F^{\m\n}\,,\quad F_{\m\n}=\partial_{\m}A_{\n}-\partial_{\n}A_\m\,,
\ee
with field equation $\partial^\m F_{\m\n}=0$, i.e. both Fronsdal equation \cite{Fronsdal:1978rb} and Maxwell-like HS equation \cite{Campoleoni:2012th} for $s=1$.
In three dimensions one can also consider Chern-Simons (CS) vector fields with free Lagrangian:
\be
\mathcal{L}^0_{s=1}=\frac12\epsilon^{\m\n\l}A_{\m}\partial_{\n}A_\l\,,
\ee
and free field equation $F_{\m\n}=0$.
It is common to call CS field ``spin one'' or ``vector'' field, but one has to be careful to not confuse it with the Maxwell field. We will mostly use the terms CS or Maxwell for the corresponding fields in the following. Since this field appears naturally in the context of HS gravity theories \cite{Prokushkin:1998bq,Gwak:2015vfb}, we study its interactions with other massless fields for completeness of our analysis.

Note that due to the difference in the free-field equations, the equivalence class that is defined for field redefinitions and gauge variations of vertices is different for CS fields. Any term that is proportional to free Maxwell field equations is obviously also on-shell trivial for CS fields. The opposite, however, is not true. The on-shell trivial cubic terms for a CS field ($i-$th field in the vertex) that are not trivial for Fierz (i.e. Fronsdal and Maxwell-like) fields are given by the expressions:
\begin{align}
G-y_i\,z_i=0\,,\quad y_i\,y_{i\pm 1}=0\,,\quad W_i=0\,,\quad V_{i\pm 1\,i}=0\,,
\quad W_{i+1}z_{i+1}=W_{i-1}z_{i-1}\,.\label{CSid}
\end{align}
Together with Fierz equations and Schouten identities, these terms define an equivalence class for cubic vertex monomials. Cubic vertices with CS fields should have trivial gauge variations in this class while not being trivial themselves.
There are three possible cases depending on the number of CS fields in the cubic vertex.

\paragraph*{Vertices with one CS field.}

A general ansatz for parity-even vertices with one CS field and two Fierz fields with spins $s_1,\,s_2$ can be written in the form:
\be
\mathcal{V}=y_1^{s_1-s_2}z_3^{s_2-1}(\a\, y_1\,z_1+\g\, y_3\,z_3)\,.
\ee
Note that $y_2\,z_2$ term is absent as it can be replaced by $y_1\,z_1$ due to identity \eqref{CSid}: $y_1\,z_1+y_2\,z_2=0$.
This vertex is gauge invariant with respect to all three gauge transformations for $\a=0$ and is not trivial only for $s_1=s_2$. It therefore defines a unique cubic vertex of  two massless fields of spin $s$ and a CS field:
\be
\boxed{\mathcal{V_{CS}}=y_3\,z_3^s\,.\label{CS}}
\ee
It is straightforward to see that this interaction corresponds to minimal coupling to vector gauge field obtained by replacing in the free action of a spin-$s$ field $\partial \to \nabla=\partial+A$, i.e. making use of covariant derivatives. This coupling has appeared for example in \cite{Gwak:2015vfb} and is obviously not applicable to Maxwell fields. The free action of a spin $s$ field, supplemented with covariant derivatives, is gauge invariant up to commutators of covariant derivatives which are proportional to the curvature of the vector field. This curvature terms are equations of motions for CS fields and can therefore be compensated by deformations of transformations for CS fields. This is not true for Maxwell fields though. The absence of minimal coupling to the electromagnetic field is known as the Velo-Zwanziger problem \cite{Velo:1970ur} and is analogous to the Aragone-Deser problem for minimal coupling to gravity. In three dimensions, the minimal coupling to gravity exists which is related to the fact that the Riemann curvature of gravity is proportional to Einstein equations in three dimensions and therefore the problematic terms can be compensated with deformations of gauge transformations of the metric (analogously to the Rarita-Schwinger coupling that leads to Supergravity). Even though the mechanisms are slightly different for spin one and spin two, in both cases the fact that the curvature tensor is on-shell trivial allows for minimal coupling. The on-shell triviality of the curvature tensor is, on the other hand, related to the absence of dynamical degrees of freedom in the bulk and opens the possibility for CS formulation.

Parity-odd CS vertices are also severely restricted. Given that the third field in the vertex is a CS field, it can for example be shown that $W_1\,y_1\,z_1=0=W_2\,y_2\,z_2$. After some algebra, it is straightforward to show that, for two Fierz fields with spins $s_1=s_2=s$, there is a two-derivative coupling to CS field with the vertex operator given by:
\be
\boxed{\mathcal{V_{CS}}^{PO}=W_1\,z_1\,z_3^{s-1}=W_2\,z_2\,z_3^{s-1}=\frac12\, U\,y_1\,y_2\,z_3^{s-2}\,.}
\ee
We skip the details of the computations here since they are elementary algebraic manipulations by straightforward application of the Schouten identities, Fierz equations and \eqref{CSid}.

\paragraph*{Vertices with two CS fields.}

In this case, the extra identities include (we assume the second and third fields are CS):
\be
y_i\,y_j=0\, (i\neq j)\,,\quad G-y_2\,z_2=0=G-y_3\,z_3\,,\quad W_2=W_3=0\,,\quad W_1\,z_1=0\,.
\ee
Using these equalities, one can easily show that there are no vertices of interactions between a massless field with spin $s$ and two CS fields if $s\geq 2$. Instead, there is a vertex of interactions of a Maxwell field and two CS fields:
\be
\boxed{\mathcal{V}_{MCS}=y_1\,z_1=-y_2\,z_2=-y_3\,z_3\,.}
\ee
For this vertex to be non-zero, the two CS fields should be charged.

There are no parity-odd vertices with two CS fields and a massless field with spin $s$.

\paragraph*{Vertices with three CS fields.}

In this case we have:
\be
W_i=0\,,\quad V_{ij}=0\,,\quad y_i\,y_j=0\,,\quad y_i\,z_i=0\,.
\ee
The only non-trivial contraction between the three fields is given by a parity-odd expression,
\be
\boxed{\mathcal{V}_{CS}^{PO}=U\,,}
\ee
which is gauge invariant and is the well-known interaction term of CS fields. For this interaction to be non-trivial, CS fields should carry non-abelian charges.

These results fit into the picture of our findings for HS fields. There are no cubic interactions between massless fields where spins do not satisfy triangle inequalities.

\section{Discussion}\label{Discussion}

In this work, we completed the program initiated in \cite{Mkrtchyan:2017ixk} providing an exhaustive classification of  covariant cubic interactions for massless bosonic fields in three dimensions. 
We found that the parity-odd cubic vertices for interactions of massless fields in three dimensions are in one-to-one correspondence with parity-even vertices. For each collection of massless fields satisfying strict triangle inequalities, there is a unique parity-odd vertex on top of the unique parity-even one \footnote{The only exception is the cubic interaction between three Maxwell fields in both parity-even and parity-odd cases. In parity-even case there are two vertices for collection of Maxwell fields --- Yang-Mills vertex and $F^3$ vertex, both requiring fully antisymmetric Chan-Paton factors. For the parity-odd case, there are three free parameters in the two-derivative vertex \eqref{PO111}, which has no definite symmetry.}. For triplet of spins  not satisfying triangle inequalities, the only cubic vertices are of ``current-interaction'' type, involving two matter fields of spin $s=0$ or $1$. 
For the triplets, that contain at least two spins greater than one, all the vertices have either two or three derivatives.

Our results should match the CFT three-point functions, as argued, e.g., in \cite{Costa:2011mg}. The uniqueness of the vertex for given triplet is in agreement with the two-dimensional CFT. The three-point functions of quasi-primaries in 2d CFT have two free parameters for every triple of spins. In our classification, for each triple we get one parity-even and one parity-odd vertex therefore match the number of independent structures. The only intriguing aspect is the missing vertices, which translate into selection rules in 2d CFT. 
Similarly to the parity-even case \cite{Mkrtchyan:2017ixk}, in the parity-odd case, missing vertices are all those containing at least two fields with spin greater than one and violating strict triangle inequalities.
Therefore, for quasi-primaries of spin values $s_1\geq s_2\geq s_3\geq 2$, all the three point functions for values $s_1\geq s_2+s_3$ are expected to be zero. This property is observed in known examples of 2d CFT's (see, e.g. \cite{Campoleoni:2011hg}), but we are not aware of a general proof.

The only massless fields that carry propagating degrees of freedom in three dimensions are scalar and Maxwell fields which are related by duality. Nevertheless, there are slight differences in vertices containing scalars and Maxwell fields observed in our classification. When we compare these vertices, one can take into account exact relation of duality between a Maxwell field $A_{\mu}$ and a scalar $\phi$, given by the relation:
\be
F_{\mu\nu}=\partial_{\mu}A_{\nu}-\partial_{\nu}A_{\mu}=\epsilon_{\mu\nu\l}\partial^{\l}\phi\,.\label{DR}
\ee
If a vertex involves the curvature of the Maxwell field, one can simply replace it with the right hand side of the equation \eqref{DR} and get a vertex for a scalar, which has opposite parity. Similarly, if the vertex contains derivative of the scalar, one can replace it with the dual of the curvature of the Maxwell field. Instead, for the vertices where one has a naked vector potential $A_\m$ this dualization is not applicable since the curl operation is not invertible and the field $A_\m$ cannot be expressed through $\phi$ locally. Let us start with parity-even vertices with Maxwell fields. There is an $(s,1,0)$ vertex, which contains $s+1$ derivatives, and in which one can dualize the Maxwell field to get parity-odd $(s,0,0)$ vertex \eqref{POs00} with $s+1$ derivatives. Alternatively, one can dualize the scalar to get the parity-odd $(s,1,1)$ vertex \eqref{POs11} with $s+1$ derivatives. Therefore, we established duality relations:
\be
\mathcal{V}_{(s,1,1)}^{PO} \leftrightarrow \mathcal{V}_{(s,1,0)}\leftrightarrow \mathcal{V}_{(s,0,0)}^{PO}\,.
\ee
This duality works for any $s\geq 1$. 

Next, there is a parity-even vertex $\mathcal{V}_{(s,1,1)}=y_1^{s-1}G$ with $s$ derivatives. Dualization of one Maxwell field leads to a vertex $\mathcal{V}_{(s,1,0)}^{PO}$ with $s$ derivatives given by \eqref{POs10}. We can further dualize the second Maxwell field and get a parity-even vertex $\mathcal{V}_{(s,0,0)}=y_1^s$ with $s$ derivatives:
\be
\mathcal{V}_{(s,1,1)}\leftrightarrow\mathcal{V}_{(s,1,0)}^{PO}\leftrightarrow \mathcal{V}_{(s,0,0)}\,.
\ee
These dualities work for $s\geq 2$, since there is no parity-odd vertex for spin configuration $(1,1,0)$, which leaves out the Yang-Mills vertex from dualization procedure. The dualization of the other cubic vertex of three Maxwell fields, $F^3$ vertex, leads to a trivial TT expression and therefore does not have a parity-odd $(1,1,0)$ dual either.

The (A)dS cubic vertices in any dimensions can be understood as deformations of flat space cubic vertices and therefore the first step towards (A)dS vertices lies in the classification of their flat counterparts.
In fact, all known Lagrangian theories with HS spectrum in three dimensions allow for flat space limit. Therefore, one may expect that the Lagrangian formulation for Prokushkin-Vasiliev theory, if existing, may also allow for a flat limit.
Even more, three-dimensional Minkowski vertices can be extended to arbitrary Einstein backgrounds due to the same reason as the absence of Aragone-Deser problem in 3d --- the obstructing terms are given by Weyl tensors and therefore vanish in three dimensions. One can even work with full non-linear gravity while constructing the action perturbatively in powers of HS fields (see \cite{Gwak:2015vfb} for such expansions of full non-linear theories). In that case, one needs to take care of the backreaction to the Einstein equations involving HS fields, which contribute to the construction of quartic and higher order vertices.

The full classification of cubic vertices is the first step towards construction of a Lagrangian for the HS theories accommodating propagating degrees of freedom which are not covered by Chern-Simons actions. Our classification is performed for the three-dimensional Minkowski background while the (A)dS extension can be considered straightforwardly. Since the main technical difficulty of (A)dS extensions is related to the non-trivial commutators of covariant derivatives, it is natural to expect that those vertices that contain many derivatives will be the most challenging. As we have seen, in three dimensions the only vertices that contain more than three derivatives are current interactions containing scalar and Maxwell fields. The AdS extensions for these vertices have already been studied in higher dimensions in \cite{Manvelyan:2009tf,Bekaert:2009ud,Mkrtchyan:2010pp}. The scalar coupling in three dimensions was studied in \cite{Prokushkin:1999xq,Ammon:2011ua,Kessel:2015kna,Lovrekovic:2018hgu}.

The main technical novelty of the three-dimensional classification provided in this work and in \cite{Mkrtchyan:2017ixk} compared to earlier work on cubic vertices in arbitrary dimensions  is related to the systematic implementation of Schouten identities in three dimensions. When considering quartic interactions of massless symmetric fields, there are relevant Schouten identities in dimensions $d\leq 7$. Therefore, the analysis of quartic order of interactions becomes more involved. We plan to address that problem both in three-dimensional and higher-dimensional contexts in the future.

We delegate some more technical discussion to appendices.
In Appendix \ref{A}, we elaborate on the possibility of writing parity-even vertices as ratios which are by themselves meaningless expressions, but can be defined and motivated {\it only in three dimensions} due to their equivalence to vertices of \cite{Mkrtchyan:2017ixk} via Schouten identities.
We study two-dimensional vertices in Appendix \ref{B}. There, the restrictions imposed by Schouten identities are much more severe and eventually allow for only vertices of the type $(s,s,1)$ and $(s,s,0)$.



\section*{Acknowledgements}

We would like to thank Andrea Campoleoni, Eduardo Conde, Dario Francia, Stefan Fredenhagen, Euihun Joung, Gabriele Lo Monaco, Stefan Theisen and Arkady Tseytlin for discussions related to the subject of this work. We are also indebted to Nicolas Boulanger for comments on the draft. The work of K.M. is supported by Alexander von Humboldt foundation. K.M. would like to thank Scuola Normale Superiore and INFN Sezione di Pisa for the hospitality extended to him during the final stage of this work. P.K. would like to thank the Albert Einstein Institute for generous support by which his modest contribution to this work became possible.

\appendix

\section{Vertices as Ratios}\label{A}
In generic dimension, the $(s_1,s_2,s_3)$ cubic interactions can be written in the basis of cubic vertices of the form\footnote{One easily checks that $s_i$ corresponds to the spins of the field $\phi_i$ by rewriting each term in $\mathcal{V}_{s_1,s_2,s_3}$ as
\begin{align}
y_1^{s_1-n}y_2^{s_2-n}y_3^{s_3-n} \sum_{\alpha+\beta+\gamma=n} c_{\alpha\beta\gamma}\, (y_1 z_1)^\alpha \,(y_2 z_2)^\beta \, (y_3 z_3)^\gamma
\end{align}
where $c_{\alpha,\beta,\gamma}$ are the trinomial coefficients. Counting the powers of say $a_1$ then gives
\begin{align}
\underbrace{s_1-n}_{y_1^{s_1-n}}+\underbrace{\alpha}_{(y_1 z_1)^\alpha \sim y_1^\alpha}+\underbrace{\beta}_{(y_2 z_2)^\beta\sim z_2^\beta}+\underbrace{\gamma}_{(y_3 z_3)^\gamma\sim z_3^\gamma} = s_1 \,.
\end{align}
}
\begin{align}
\mathcal{V}_{s_1,s_2,s_3} = \sum_{n=0}^{\text{min}(s_1,s_2,s_3)} g_n \; y_1^{s_1-n}y_2^{s_2-n}y_3^{s_3-n} G^n \,,\label{MB}
\end{align}
where $g_n$ are undetermined constants. The $n-$th term in the sum has $s_1+s_2+s_3-2n$ derivatives. Therefore, the term with the minimal number of derivatives is $n=\text{min}(s_1,s_2,s_3)$ while $n=0$ contains the maximal number of derivatives. These two bounds are commonly referred to as the lower and upper Metsaev bound respectively.

For a given dimension, one may be able to construct vertices which violate these bounds due to the presence of Schouten identities. As an example, one can construct vertices corresponding to minimal $(s,s,2)$ coupling to gravity in three dimensions. These vertices only involve two derivatives and therefore violate the lower Metsaev bound for $s>2$. One may try to express these vertices in the form of \eqref{MB}\footnote{Similar attempt has been made for four-dimensional vertices in \cite{Sleight:2016xqq}.}.

We start from
\begin{align}
\label{eq:spin2minimal}
\mathcal{V} = \frac{G^s}{y_3^{s-2}} \,,
\end{align}
which by itself is of course a nonsensical expression. However, as we will see shortly, one can make sense of it \textit{only in three dimensions} using Schouten identities. Note that this vertex only contains two derivatives and obviously fulfills $D_i \mathcal{V} = 0$. By using the definition of $G$ one obtains
\begin{align*}
\frac{G^s}{y_3^{s-2}} &= \sum_{k=0}^s \binom{s}{k} \, \frac{(G-y_3 z_3)^k (y_3 z_3)^{s-k}}{y_3^{s-2}} \\
&= \underbrace{z_3^s y_3^2 + s (G-y_3 z_3) \, y_3 z_3^{s-1}}_{\mathcal{V}_{min}} + \underbrace{(G-y_3 z_3)^2}_{\Delta} \underbrace{\sum_{k=0}^{s-2} \binom{s}{k+2} \frac{(G-y_3 z_3)^k z_3^{s-k+2}}{y_3^k}}_{R} \,. 
\end{align*}
The last term however vanishes due to the Schouten Identity \eqref{eq:twoDerivDDI}. Therefore, we have constructed a gauge invariant $(s,s,2)$ TT vertex involving only two derivatives as can be seen as follows
\begin{align}
0 = D_i \mathcal{V} =  D_i \left( \mathcal{V}_{min} + \Delta \times R  \right) = D_i \mathcal{V}_{min} + D_i(\Delta) \times R + \Delta \times D_i(R).
\end{align}
Since $D_i (\text{Schouten Identities}) \subset \text{Schouten Identities}$, it then follows that $D_i \mathcal{V}_{min} = 0$.

This procedure can fail in subtle ways. To illustrate this, let us consider one-derivative minimal coupling to Maxwell field which is defined, similarly to \eqref{eq:spin2minimal}, through a ratio:
\begin{align}
\frac{G^s}{y_3^{s-1}}&= \sum_{k=0}^s \binom{s}{k} \, \frac{(G-y_3 z_3)^k (y_3 z_3)^{s-k}}{y_3^{s-1}} \\
&= \underbrace{z_3^s y_3 + s (G-y_3 z_3) \, z_3^{s-1}}_{\mathcal{V}_{min}} + \underbrace{(G-y_3 z_3)^2}_{\Delta} \underbrace{\frac{1}{y_3}\sum_{k=0}^{s-2} \binom{s}{k+2} \frac{(G-y_3 z_3)^k z_3^{s-k+2}}{y_3^k}}_{R} \,.  \,,\label{eq:spin1minimal}
\end{align}
In this case,
\begin{align}
\Delta = (G-y_3 z_3)^2\,, && R=\frac1{y_3} \, \sum_{k=0}^{s-2} \binom{s}{k+2} \frac{(G-y_3 z_3)^k z_3^{s-k+2}}{y_3^k}\,,\label{DR1}
\end{align}
it follows for example that
\begin{align}
D_2(\Delta) \times R \sim (G-y_3 z_3) y_2 y_3 \times \frac1{y_3} \, \sum_{k=0}^{s-2} \binom{s}{k+2} \frac{(G-y_3 z_3)^k z_3^{s-k+2}}{y_3^k}\,,
\end{align}
and due to the pole in $y_3$ the first term in the sum is no longer proportional to a Schouten identity after canceling terms common to the denominator and numerator. Note that such a pole would not arise for the case of minimal coupling to gravity \eqref{eq:spin2minimal}.
It is interesting to notice that the same ratio \eqref{eq:spin1minimal} defines a minimal coupling to Chern-Simons field given in \eqref{CS}:
\be
\mathcal{V}_{min}=y_3\,z_3^s\,.
\ee
This is due to first identity of \eqref{CSid} which allows to replace \eqref{DR} by
\be
\Delta = G-y_3\,z_3=0\,,\quad  R= \, \sum_{k=0}^{s-1} \binom{s}{k+1} \frac{(G-y_3 z_3)^k z_3^{s-k+2}}{y_3^k}\,,
\ee
and extending the argument given for spin two minimal coupling to this case. As we have seen, the schematic way of writing the vertex as ratios works consistently only for vertices that are otherwise shown to exist in covariant formulation due to Schouten identities. We conclude that this rewriting is just a curiosity and does not provide with any new insights.

\section{d=2}\label{B}

In two dimensions the Schouten identities can be used to eliminate the d'Alambertian term in the Fronsdal action for $s\geq 2$ in favour of trace and divergence terms, which renders the free theory to be trivial for TT fields.
For example, the spin two Fronsdal equation itself is proportional to Schouten identity. This is a `linearization' of the statement that Einstein-Hilbert action is topological and there are no Einstein equations of motion for the metric in 2d.
For $s>2$, the massless equation \eqref{Fierz1} is a consequence of the two other equations \eqref{Fierz2} and \eqref{Fierz3}. One may study massless HS fields in fully reducible Maxwell-like formulation in $d=2$. There, each even rank field carries a single scalar mode.

HS theories in 2d are making use of BF-type actions \cite{Alkalaev:2013fsa}. Nevertheless, if one insists on Fronsdal formulation and tries to derive TT cubic couplings, following observations are in order. Some of the Schouten identities that can be derived in this case are:
\ba
y_1\,y_2\,y_3=0\,,\quad y_i^2\, y_j=0\,, \quad y_i\,y_j\,z_j=0\,,\quad G^2=0\,,\\
 z_i\,(G-y_{i\pm 1}z_{i\pm 1})=0\,, \quad z_1\,z_2\,z_3=0\,.
\ea
These identities imply that any term with more than two derivatives is TT trivial and there are no candidate TT expressions for $s_1\geq s_2\geq s_3\geq 2$. The only parity-even TT vertices that can be written down, necessarily have scalar or Maxwell fields involved and are given by $(s,s,0)$ vertex $\mathcal{V}_{(s,s,0)}=y_1y_2z_3^{s-1}$, $(s,s,1)$ vertex $\mathcal{V}_{(s,s,1)}=y_3z_3^s$ ($s\geq 2,\, s=0$) and usual Yang-Mills $(1,1,1)$ vertex $\mathcal{V}_{(1,1,1)}=G$. These observations may be useful in the attempts to construct HS gravity theories in two dimensions.


\end{document}